# Quantum Oscillations in Black Phosphorus Two-dimensional Electron Gas

Likai Li[1,4], Guo Jun Ye[2,3,4], Vy Tran[5], Ruixiang Fei[5], Guorui Chen[1,4], Huichao Wang[6], Jian Wang[6], Kenji Watanabe[7], Takashi Taniguchi[7], Li Yang[5], Xian Hui Chen[2,3,4*] and Yuanbo Zhang[1,4*]

[1]*State Key Laboratory of Surface Physics and Department of Physics, Fudan University, Shanghai 200438, China*

[2]*Hefei National Laboratory for Physical Science at Microscale and Department of Physics, University of Science and Technology of China, Hefei, Anhui 230026, China*

[3]*Key Laboratory of Strongly Coupled Quantum Matter Physics, University of Science and Technology of China, Hefei, Anhui 230026, China*

[4]*Collaborative Innovation Center of Advanced Microstructures, Nanjing University, Nanjing, 210093, China*

[5]*Department of Physics and Institute of Materials Science and Engineering , Washington University in St. Louis, St. Louis, MO 63136, USA*

[6]*International Center for Quantum Materials, School of Physics, Peking University, Beijing 100871, China*

[7]*Advanced Materials Laboratory, National Institute for Materials Science, 1-1 Namiki, Tsukuba, 305-0044, Japan.*

*Email: zhyb@fudan.edu.cn, chenxh@ustc.edu.cn

**Two-dimensional electron gases (2DEG), have been an important source of experimental discovery[1,2] and conceptual development in condensed matter physics for decades[3]. When combined with the unique electronic properties of two-dimensional crystals, rich new physical phenomena can be probed at the quantum level[4,5]. Here, we create a new 2DEG in black phosphorus – a recent member of the two-dimensional (2D) atomic crystal family[6–8] – using a gate electric field. We achieve high carrier mobility in black phosphorus 2DEG by placing it on a hexagonal boron nitride (h-BN) substrate. This allows us, for the first time, to observe quantum oscillations in this material. The temperature and magnetic field dependence of the oscillations yields crucial information about the system, such as cyclotron mass and lifetime of its charge carriers. Our results, coupled with the fact that black phosphorus possesses anisotropic energy bands with a tunable, direct bandgap[6–15], distinguishes black phosphorus 2DEG as a novel system with unique electronic and optoelectronic properties.**

2DEGs form when electrons are tightly confined to a quantum well, as typically found at the 2D heterojunction in a field-effect transistor (FET). The recent emergence of 2D layered materials has greatly enriched the prospective host materials of 2DEGs. These atomically thin crystals provide a maximum level of quantum confinement, while introducing remarkable new physics to the 2DEG due to their distinctive electronic structures[4,5,16]. The charge carrier mobility is an important characteristic of 2DEGs – a high mobility is required to observe physical phenomena that are unique to 2D systems such as the integer and fractional quantum Hall effects[1]. Furthermore, the mobility plays a crucial role in the application of 2DEGs in high speed devices[18], which are ubiquitous in consumer electronics. However, despite the large number of semiconductors currently available, 2DEGs with high mobility (> 1000 cm$^2$/Vs) are only found in a selected few.

In this letter we report a new type of high mobility 2DEG that we induce on the surface of black phosphorus using a gate electric field. A new member of the 2D atomic crystal family, black phosphorus is a stable allotrope of phosphorus that features an anisotropic puckered honeycomb lattice in each layer. Bulk black phosphorus is a semiconductor with a direct bandgap of ~ 0.3 eV (Ref. 19–23), which is expected to progressively increase to ~2.0 eV as the crystal is thinned down to monolayer (also referred to as phosphorene)[9,10]. Field-effect transistors (FETs) made out of few-layer phosphorene has recently been shown to exhibit high conductance modulation[6–8,23,24], anisotropic charge transport[7,13,15,23], and high carrier mobility[6–8,23,24]. Here we achieve a Hall mobility in our few-layer phosphorene 2DEG that is one order of magnitude higher than the previous record. This high mobility allowed us for the first time to observe the quantum oscillations in black phosphorus, and we were able to probe the formation of Landau levels (LL) in our few-layer phosphorene 2DEG in the extreme quantum limit.

We apply the FET principle to black phosphorus, and our device structure is shown in Fig. 1a and 1b. The key to the high mobility in our sample is two-fold: black phosphorus crystals with much improved quality (see Methods); and h-BN substrate, which has been demonstrated to preserve the ultra-high mobility in graphene samples[17].

We start with black phosphorus and h-BN thin flakes that are freshly cleaved on Polydimethylsiloxane (PDMS) film and Si wafer covered with 286 nm $SiO_2$, respectively. The black phosphorus flake on PDMS is then transferred onto the h-BN supported on $SiO_2$/Si wafer following Ref. 25, and the dry transfer method ensures an extremely clean interface between the two flakes. To avoid the degradation of black phosphorus surface, the mechanical exfoliation and transfer process were performed in an Ar atmosphere with $O_2$ and $H_2O$ content kept under 1 ppm. We then fabricate electrodes on top using standard electron beam lithography, followed by metal deposition (Au/Cr, 60 nm and 2 nm, respectively). The degenerately doped Si serves as the back gate that we use to induce electron or hole carriers in the sample. The thickness of black phosphorus flakes were chosen to be ~ 10 nm and we found that slight variation in the sample thickness does not affect our experimental results.

We observe drastically improved charge carrier mobility in our black phosphorus FETs on h-BN substrate at low temperatures. Here two types of mobility are obtained on the same sample for comparison: a) the field-effect mobility $\mu_{FE}$, extracted from the line fit of the transfer characteristics (Fig. 1c), and b) Hall mobility $\mu_H = \sigma/n_h e$, where $\sigma$ is the sheet conductance, $n_h$ is the sheet carrier density determined by Hall measurement, and $e$ is the charge of an electron. As shown in Fig. 1c, the room temperature $\mu_{FE}$ for holes and electrons is 400 cm²/Vs and 83 cm²/Vs, respectively. These field-effect mobility values are comparable to those previously obtained in black phosphorus FETs on $SiO_2$ substrate[6,7]. As the temperature $T$ is lowered, however, $\mu_{FE}$ increases by one order of magnitude, reaching record-high field-effect mobility for both holes (3900 cm²/Vs) and electrons (1600 cm²/Vs) at $T = 1.5$ K. The Hall mobility follows a similar trend as the temperature is lowered: $\mu_H$ starts at modest values at room temperature, increases continuously till $T \sim 30$ K, and levels off to high mobility values (~ 2000 cm²/Vs for holes and ~ 900 cm²/Vs for electrons) at low temperatures. The general behavior of the temperature-dependent $\mu_H$ observed here is typical of semiconducting devices[26], which indicates that the high-temperature ($T > 30$ K) and low-temperature ($T < 30$ K) mobility is dominated by phonon and impurity scattering, respectively. The transition temperature between the two regimes is much lower than that reported in black phosphorus FETs fabricated on SiO₂ substrate ($T \sim 100$ K, Ref. 6, 23). Coupled with the fact that our low-temperature Hall mobilities are one order of magnitude higher than previously reported values[6,23], our results indicate significantly reduced impurities in our samples.

The considerably improved mobility allows us to observe the Shubnikov-de Haas (SdH) oscillations in the magneto-resistance $R_{xx}$ of our sample – the first such observation in black phosphorus. Fig. 2a displays the $R_{xx}$ as a function of magnetic field $B$ at varying gate voltages $V_g$, and pronounced oscillations are seen at both hole doping ($V_g < 0$) and electron doping ($V_g > 0$). The oscillations start at a critical magnetic field $B_c \sim 8$ T at the hole side and a $B_c \sim 15$ T at the electron side. Such critical fields correspond to the minimum magnetic field needed for the charge carriers to complete a cyclotron orbit before getting scattered, from which an estimation of the carrier mobility can be obtained: $\mu \sim 1/B_c$. Our observed critical fields yields a hole

Page **5** of **19**

mobility of ~1200 cm²/Vs and an electron mobility of ~ 600 cm²/Vs. These estimations agree with the Hall mobility values reasonably well, and further confirm the high quality of our sample. Quantum oscillations are also observed when we sweep $V_g$ while keeping the magnetic field fixed at $B = 31$ T (Fig. 2b). We identify the LL index of each peak by calculating its corresponding filling factor, i.e. the number of LLs (each containing a carrier density of $2eB/h$, where $h$ is the Plank constant and the factor of 2 here is from spin degeneracy[27]) filled by the carriers induced by the gate, $n_g = C_g V_g / e$. Here $C_g = 1.17$ Fcm⁻² is the capacitance per unit area for the back gate with 286 nm SiO₂ and 16 nm h-BN as the dielectric. The results show that we are approaching the lowest LL and the transport is already in the extreme quantum limit (Fig. 2b). In addition, the splitting of every LL into two means that the spin degeneracy is fully lifted at $B = 31$ T, suggesting a potentially large Landé *g*-factor for the carriers.

The constant period of the oscillations seen in Fig. 2b is *prima facie* evidence of two dimensionality of the charge transport in our sample. We attribute the tight confinement of the out-of-plane motion of the carrier to the narrow quantum well at the interface induced by the gate. The width of the quantum well is determined by *ab-initio* calculations to be $\lambda \sim 2$ nm (see Supplementary Information, and also Ref. 13). Our calculation further shows that doped free carriers are mostly confined within ~ 2 atomic layers under our gate voltages, as shown in Fig. 2e.

Indeed, additional evidence of the two dimensionality of our system is uncovered by an analysis of the gate-dependence of the SdH oscillations shown in Fig. 2a. The oscillations are described semi-classically by[28]:

$$\Delta R_{xx} = R(B,T)\cos[2\pi(B_F/B+1/2)] \qquad (1)$$

Where $R(B,T)$ is the SdH oscillation amplitude and $B_F$ is the oscillation frequency in $1/B$. We obtain $B_F$ from the slope of the Landau fan diagram, in which the sequence of values of $1/B_n$ of the $n$th minimum in $R_{xx}$ are plotted against their corresponding LL index $n$, for each applied gate voltage (Fig. 2c). (Here the Zeeman splitting of the SdH oscillations is neglected, and we only pick the main dips in the magneto-resistance to plot the fan diagrams.) Measurement of $B_F$ yields the sheet density of carriers participating in the SdH oscillations, $n_s$, through $n_s = 2eB_F/h$ (Ref. 27), which are plotted as a function of gate voltage in Fig. 2d (red circles). The excellent match between $n_s$ and the gate-induced carrier density (Fig. 2d, black lines) means that all the gate-induced carriers participate in the single-frequency SdH oscillations at each gate voltage. This observation indicates that no other bands are involved in the transport, and all carriers in our black phosphorus 2DEG resides in the lowest subband of the quantum well – a result confirmed by our calculations (Supplementary Information). We note that the linear fit to the Landau fan diagrams on the electron side intercepts the $n$-index axis at ~ 0.5. This non-zero intercept is not manifestation of non-trivial Berry's phase[30] (there is no such phase in energy bands of pristine black phosphorus[31]), but rather a result of spin splitting of the Landau levels, which we shall discuss later.

We further probe the 2D nature of the electron gas by examining the shape of its Fermi surface (FS). The external cross-section of the FS $S_F$, as seen by a magnetic field applied perpendicular to the sample plane, is in general related to the SdH oscillation frequency by $B_F = hS_F/(2\pi)^2 e$ (Ref. 28). As the sample normal is tilted away from the field by an angle $\alpha$ (Fig. 3a, inset), the $S_F$ seen by the field is modified by a factor that's determined by the shape of the FS and $\alpha$. We can therefore extract important information on the FS by measuring $B_F$ as a function of $\alpha$. For a 2D FS the factor is simply $\cos(\alpha)$, so $B_F \propto 1/\cos(\alpha)$. This angular dependence is exactly what we have observed in our sample for $\alpha$ up to 43° (Fig. 3b, red circles). (For $43° < \alpha < 90°$, the oscillations are not resolved.) We also tilted the sample about a second in-plane axis that is 90° apart, and the angular dependence remains the same (Supplementary Information). For comparison, we have plotted the expected angle dependence of $B_F$ in Fig. 3b if the three-dimensional (3D) FS of bulk black phosphorus is assumed. (Here the angle dependence depends on the crystal orientation of our sample, which we determine by Raman spectroscopy. See supplementary for details). The drastic departure from our observation provide unambiguous proof that the electron gas in our sample is 2D.

Having established the 2D nature of our system, we now extract more specific information on the black phosphorus 2DEG. We do so by analyzing the amplitude of the SdH oscillations $R(B,T)$, which can be generally decomposed into three reduction factors[28]:

$$R(B,T) = R_T R_D R_S \tag{2}$$

These factors contain crucial information on the cyclotron effective mass, lifetime and spin of the 2D charge carriers, which we discuss below.

The $T$ dependence of the amplitude is described by the first reduction factor $R_T \propto \lambda(T)/\sinh\lambda(T)$, and the thermal factor $\lambda(T)$ is given by $\lambda(T) = 2\pi^2 k_B T m^*/\hbar eB$. Here $k_B$ is the Boltzmann constant and $m^*$ is the cyclotron mass of the carriers. By fitting the observed thermal damping of the oscillations with $R_T$ (Fig. 4a and 4b, $V_g$ is fixed at -80 V), we are able to extract the cyclotron mass of holes $m_h^* = 0.34 m_0$, where $m_0$ is the mass of a bare electron (Fig. 4c). A similar analysis on the electron side ($V_g = +100$ V) yields an electron cyclotron mass of $m_e^* = 0.47 m_0$ (Fig. 4d, see Supplementary Information for detailed analysis). Both the hole and electron cyclotron mass is significantly larger than the values obtained from previous spectroscopic measurement[32] and *ab initio* calculations (Supplementary Information) performed on the bulk material. We attribute the abnormally large $m^*$ to the 2D nature of our electron gas – as the electrons are confined to the 2 atomic layers at the surface, the band dispersion changes appreciably, giving rise to larger $m^*$ for both holes and electrons. First-principle hybrid functional calculations for a two-layer system indeed gives the cyclotron $m^*$ values for electrons (0.43 $m_e$) and for holes (0.48 $m_e$), which are in better agreement with our experimental results, even though difference still exists due to difficulties in fitting the strongly anisotropic band dispersion (Supplementary Information).

The carrier lifetime $\tau$ of the black phosphorus 2DEG can be obtained by analyzing the second reduction factor $R_D = e^{-D}$ in Eq. (2), which is also referred to as Dingle factor. Here $D = \pi m^* \Gamma / eB$, with $\Gamma = 1/\tau$ the scattering rate. Because $R(B,T) \propto e^{-D}\lambda(T)/\sinh\lambda(T)$, we may find $\Gamma$ from the slope in the semi-log plot of $R(B,T) \sinh \lambda(T) B$ as a function of $1/B$ (Fig. 4e). We find that $\tau = 0.11 \, \text{ps}$ for holes and $\tau = 0.12 \, \text{ps}$ for electrons. These carrier lifetimes translate to a mobility of 550 cm²/Vs for holes and 450 cm²/Vs for electrons through the relation $\mu = e\tau/m^*$. These values are lower than the Hall mobilities obtained earlier, because $\tau$ is in principle smaller than the transport lifetime (which determines the Hall mobility).

Finally, we investigate effects of spin that is captured in the third reduction factor $R_S = \cos\phi$ in Eq. (2), where $\phi$ is the phase difference of spin-up and spin-down LLs introduced by Zeeman splitting[28]. The phase difference is directly proportional to the ratio between two energy scales: Zeeman energy $\varepsilon = g\mu_B B$ and the cyclotron energy $\hbar\omega_c = eB/m^*$. Here $\mu_B = e\hbar/2m_0$ is Bohr magneton, so $\phi$ takes the form $\phi = \pi g m^*/2m_0$, which does not depend on temperature or magnetic field. The lack of $T$ or $B$ dependence makes a quantitative measurement of $g$ difficult in our experiment. $R_S$, however, is still of fundamental importance by providing a global phase of 0 or $\pi$ to the SdH oscillations depending on its sign. Such a global phase is easily identified from the intercept in the Landau fan diagram, with 0 and 0.5 corresponding to a phase of 0 and $\pi$ (and thus a positive and negative $R_S$), respectively. Our experimental results, shown in Fig. 2d, unambiguously point to a positive $R_S$ for holes and a negative $R_S$ for electrons. It then follows that $\varepsilon < \hbar\omega_c/2$ for holes and $\varepsilon > \hbar\omega_c/2$ for electrons, which provide an upper bound ($g < 2.94$) and a lower bound ($g > 2.13$) for hole and electron g-factor, respectively. The relative energy spacing of spin-split LLs inferred from our experiment is illustrated in Fig. 4e. We note that for electrons the Zeeman gap is larger than the inter-LL gap in a magnetic field (the opposite is true for holes). This is indeed corroborated by our observation that the dips at the Zeeman gap are more prominent than those at the inter-LL gap for electrons (Fig. 2a and 2b).

To summarize, we have successfully created a field-induced high mobility 2DEG in black phosphorus in an FET structure. The high quality of the 2DEG allows us to observe SdH oscillation for the first time in black phosphorus. We determine the cyclotron mass of both holes (0.34 $m_0$) and electrons (0.47 $m_0$), which are notably larger than those found in bulk crystal, possibly due to the 2D nature of the electron gases. We also obtain the carrier lifetime ($\tau \sim 0.1 \text{ps}$), and set the upper bound ($g < 2.94$) and lower bound ($g > 2.13$) for the Landé $g$-factor of holes and electrons, respectively. Our experimental results introduce black phosphorus 2DEGs to the elite family of high quality electron systems, and open the door to future research on quantum transport in black phosphorus.

**Methods**

**Sample Growth.** Black phosphorus bulk crystal was synthesized under a constant pressure of 1 GPa. The high-pressure environment was provided by a cubic-anvil-type apparatus (Riken CAP-07). We heated red phosphorus (Aladdin Industrial Corporation, 99.999% metals basis) to 1,000 °C, stayed for one hour at 1,000 °C, and then slowly cooled down to 500 °C at a cooling rate of 100 °C per hour. Asymmetric graphite heater was used to create a constant temperature gradient across the growth chamber. During cooling process the temperature gradient prevents crystallization nuclei from forming at random locations, and helps concentrate the nuclei at the cooler end of the chamber. The asymmetric heater therefore effectively mitigates the competition of crystal growth direction, and improves crystal quality.

**Acknowledgements**

We thank Feng Wang, Xi Lin, Yan-Wen Tan, Lanpo He and Yanwen Liu for helpful discussions, Jun Zhao, Qisi Wang and Yao Shen for help with sample preparation, and Scott Hannahs, Tim Murphy, David Graf, Jonathan Billings, Bobby Pullum, Luis Balicas and Bing Zeng for help with measurements in high magnetic fields. A portion of this work was performed at the National High Magnetic Field Laboratory, which is supported by National Science Foundation Cooperative Agreement No. DMR-1157490, the State of Florida, and the U.S. Department of Energy. Part of the sample fabrication was conducted at Fudan Nano-fabrication Lab. L.L., G.C. and Y.Z. acknowledge financial support from the National Basic Research Program of China (973 Program) under the grant Nos. 2011CB921802 and 2013CB921902, and NSF of China under the grant No. 11034001. G.J.Y and X.H.C. acknowledge support from the 'Strategic Priority Research Program' of the Chinese Academy of Sciences under the grant No. XDB04040100 and the National Basic Research Program of China (973 Program) under the grant No. 2012CB922002. V.T., R.F. and L.Y. are supported by NSF DMR-1207141. H.W. and J.W. are supported by National Basic Research Program of China under the grant No. 2013CB934600, and NSF of China under the grant No. 11222434.
**Author contributions**

L.L. fabricated black phosphorus devices, performed transport measurements, and analyzed the data. G.J.Y. and X.H.C. grew bulk black phosphorus crystals. V.T., R.F.

Page **16** of **19**

and L.Y. did theoretical calculations. G.C., H.W. and J.W. helped with the transport measurement. K.W. and T. T. grew bulk h-BN. Y.Z. and X.H.C. co-supervised the project. L.L. and Y.Z. wrote the paper with input from all authors.

**Figure captions**

**Figure 1 | Black phosphorus FET and its basic characteristics. a,** Optical image of black phosphorus FET with h-BN substrate. **b,** Schematic cross-sectional view of the FET structure. **c,** Sample conductance measured as a function of gate voltage at $T=1.5$ K (black) and $T=300$ K (red). Field-effect mobility $\mu_{FE}$ is obtained from line fit to the linear part of the conductance (broken lines) for both holes ($V_g < 0$) and electrons ($V_g > 0$). At $T=300$ K, $\mu_{FE}$ of holes and electrons are found to be 400 cm$^2$/Vs and 83 cm$^2$/Vs, respectively. As $T$ is lowered to 1.5 K, $\mu_{FE}$ reaches 3900 cm$^2$/Vs for holes and 1600 cm$^2$/Vs for electrons. **d,** Hall mobility $\mu_H$ of holes (green) and electrons (blue) as a function of temperature.

**Figure 2 | SdH oscillations in black phosphorus 2DEG a,** $R_{xx}$ as a function of magnetic field measured at varying gate voltages. SdH oscillations are observed for both holes (upper panel) and electrons (lower panel). **b,** $R_{xx}$ (black) and $R_{xy}$ (red) measured as a function of $V_g$ with magnetic field fixed at $B=31$ T. LL index of each peak is marked. Spin degeneracy is lifted at each LL and the arrows indicate the spin-up and spin-down LLs. **c,** Landau diagrams for SdH oscillations at different gate voltages. LLs of holes are shown with positive indices, and those of electrons are shown

with negative indices. The lines are linear fit to the data, from which we extract $B_F$ as the slope. The *n*-axis intercept is 0 for holes, and 0.5 for electrons. **d,** Carrier density obtained from $B_F$ extracted from **c**. Lines indicate carrier density expected from gate-induced doping. **e,** Calculated bulk carrier distribution $n_{3D}$ at the surface of black phosphorus at varying $V_g$. The free carriers are mainly confined to ~ 2 atomic layers at $V_g < -70$ V. See Supplementary Information for details of the calculation.

**Figure 3 | Angular dependence of SdH oscillations in black phosphorus 2DEG. a,** $R_{xx}$ as a function of magnetic field measured at varying tilt angle $\alpha$. Value of $\alpha$ is shown next to each curve. Inset: measurement configuration. *x* and *y* axis indicate the crystal orientation of the sample, which is obtain by angle-dependent Raman spectroscopy (Supplementary Information). Red dashed line marks the tilt axis. **b,** $B_F$ extracted from **c** plotted as a function of $1/\cos(\alpha)$. Data are well fitted with a straight line, indicating the 2D nature of the black phosphorus 2DEG. Black curve is the angular dependence of $B_F$ expected from the 3D Fermi surface of bulk black phosphorus.

**Figure 4 | Cyclotron mass, carrier lifetime and Zeeman splitting in black phosphorus 2DEG. a,** $R_{xx}$ as a function of $1/B$ measured at different temperatures with $V_g$ fixed at -80 V. No Zeeman splitting is observed at $1/B > 0.04$ T$^{-1}$ ($B < 25$ T), and $R_{xx}$ exhibits a single SdH oscillation frequency, which is confirmed by the single peak in the Fourier transform shown in the inset. **b,** SdH oscillation amplitude extracted from **a** as a function of temperature at different magnetic field. Solids lines

are fit by temperature reduction factor $R_T$ (defined in the text). **c,** Hole cyclotron mass (left panel) and electron cyclotron mass (right panel) obtained from the fitting of temperature-dependent of oscillation amplitude shown in **b**. Both hole and electron cyclotron masses are larger than values obtained from previous experiment (black lines)[32] on bulk crystal. Red lines are our *ab initio* calculations on a bi-layer structure (Supplementary Information). **d.** Dingle plots of $\ln[\Delta R(B,T)\sinh\lambda(T)B]$ versus $1/B$ for holes (upper panel) and electrons (lower panel). A lifetime of $\tau \sim 0.1$ ps is obtained for both types of carriers from line fit of the data (solid lines) **e.** Schematic energy level diagram of holes (left) and electrons (right) showing the Zeeman splitting of the LLs. Our analysis indicates that $\varepsilon < \hbar\omega_c/2$ for holes and $\varepsilon > \hbar\omega_c/2$ for electrons (see text).

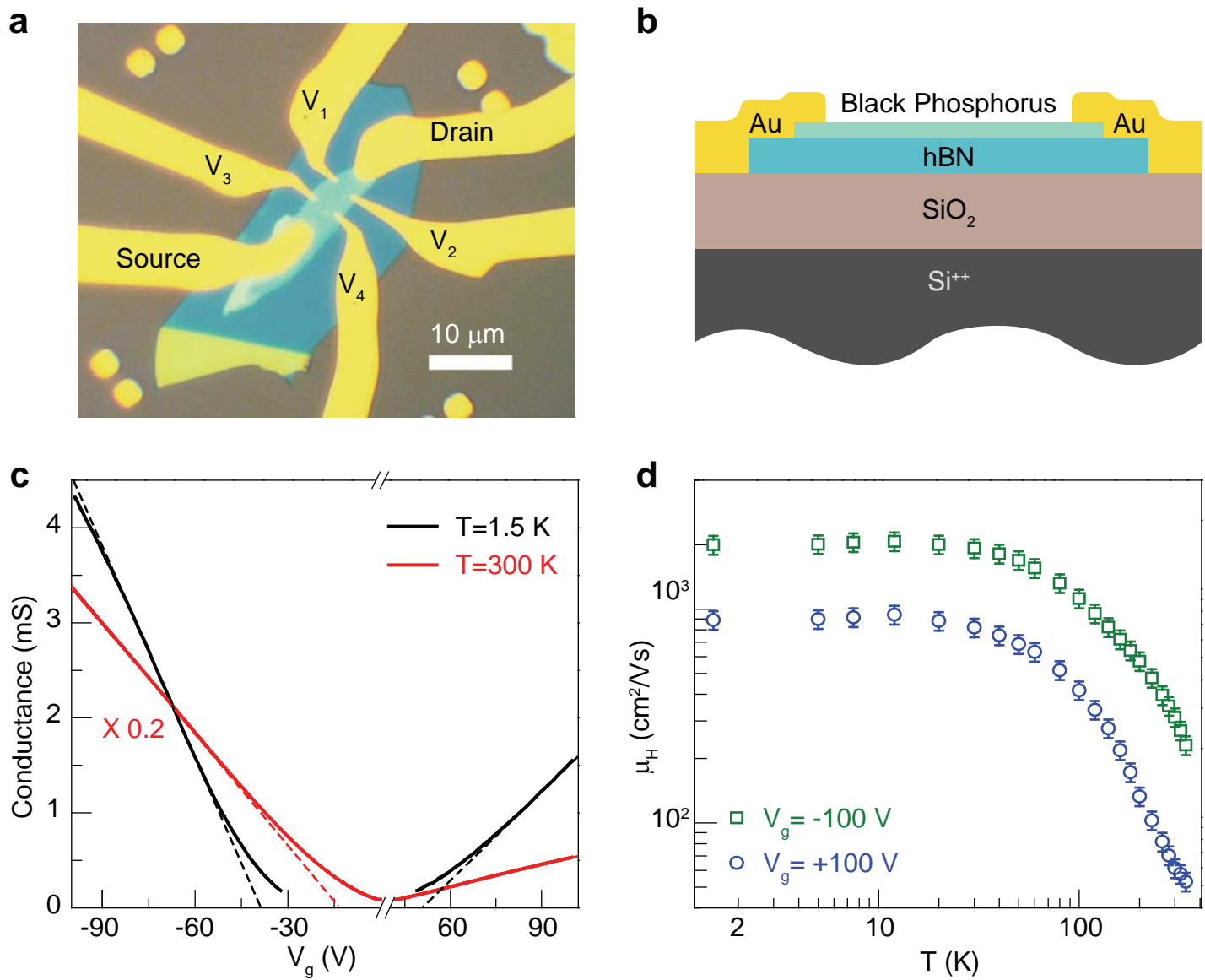

Likai Li *et al.*   Figure 1

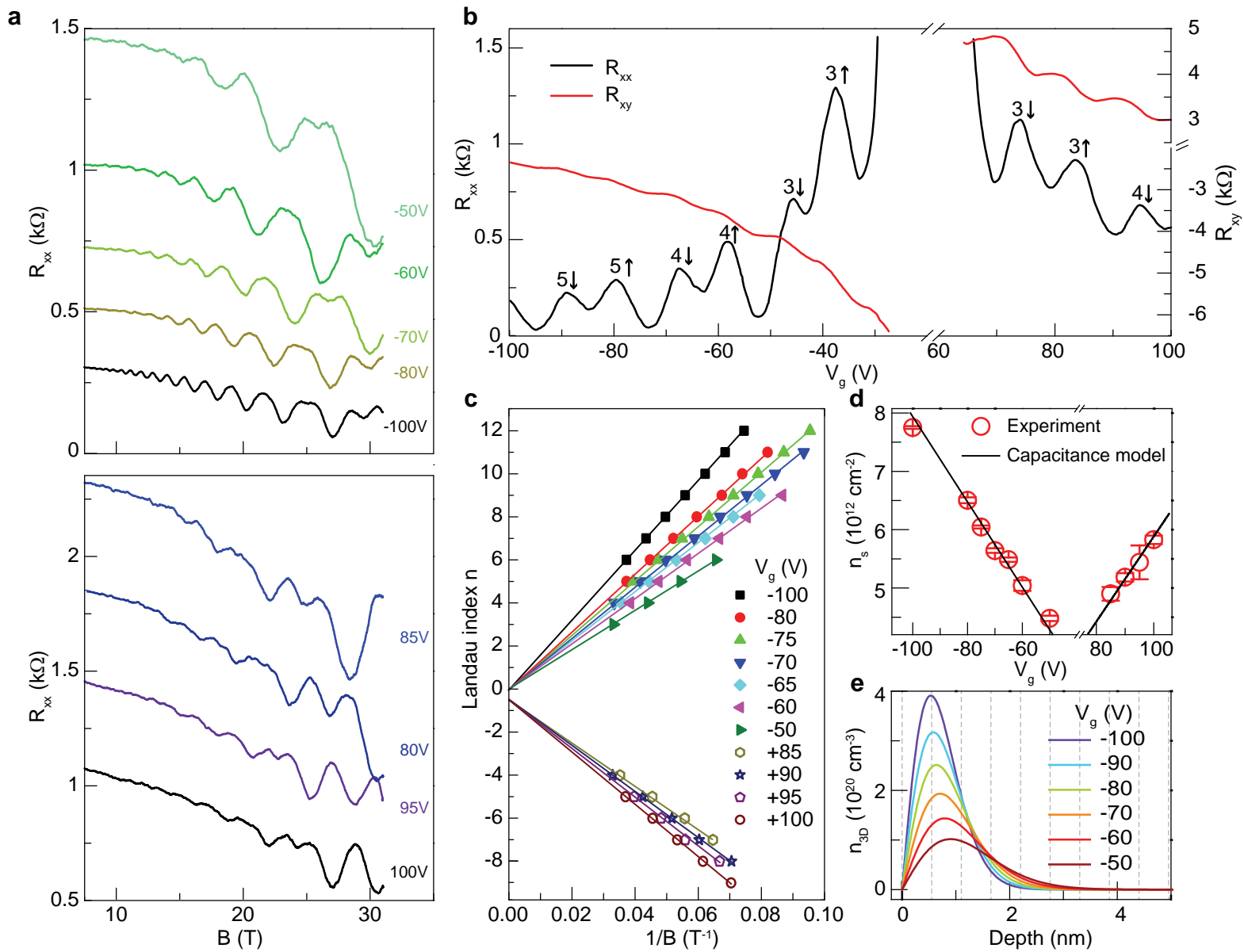

Likai Li *et al.* Figure 2

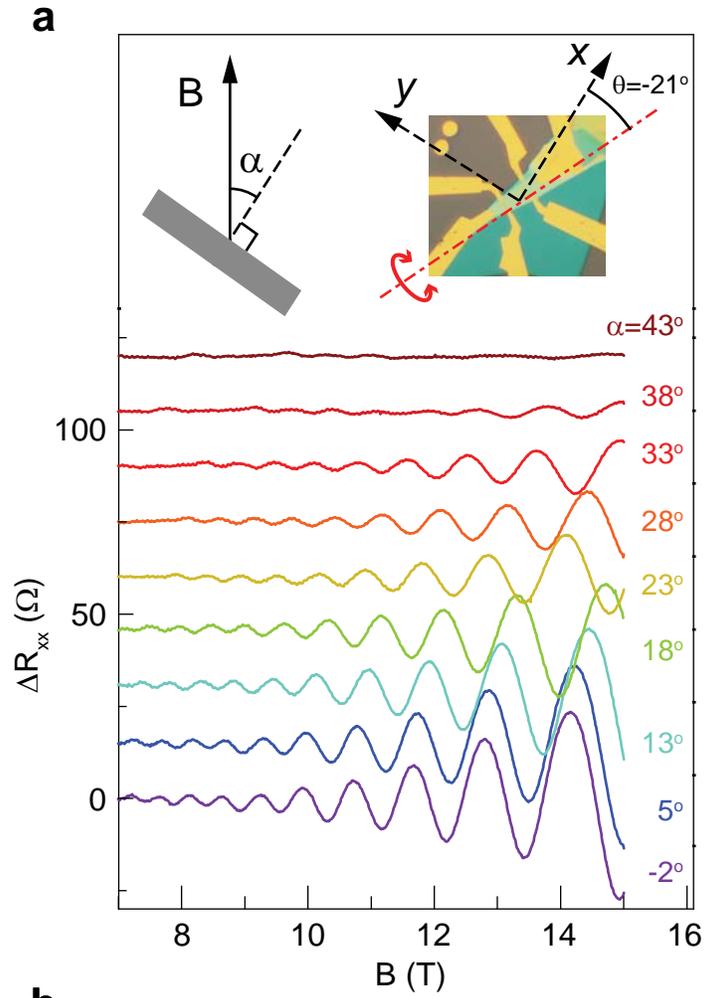
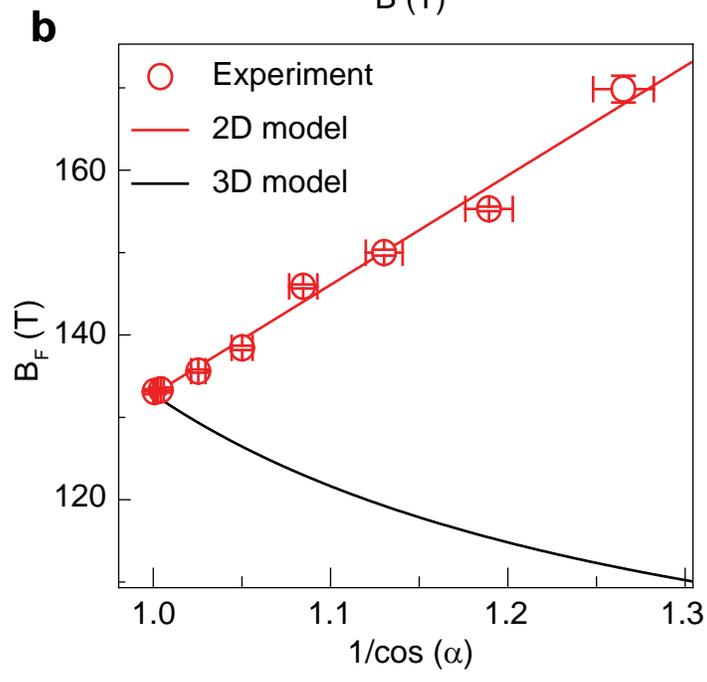

Likai Li *et al.*   Figure 3

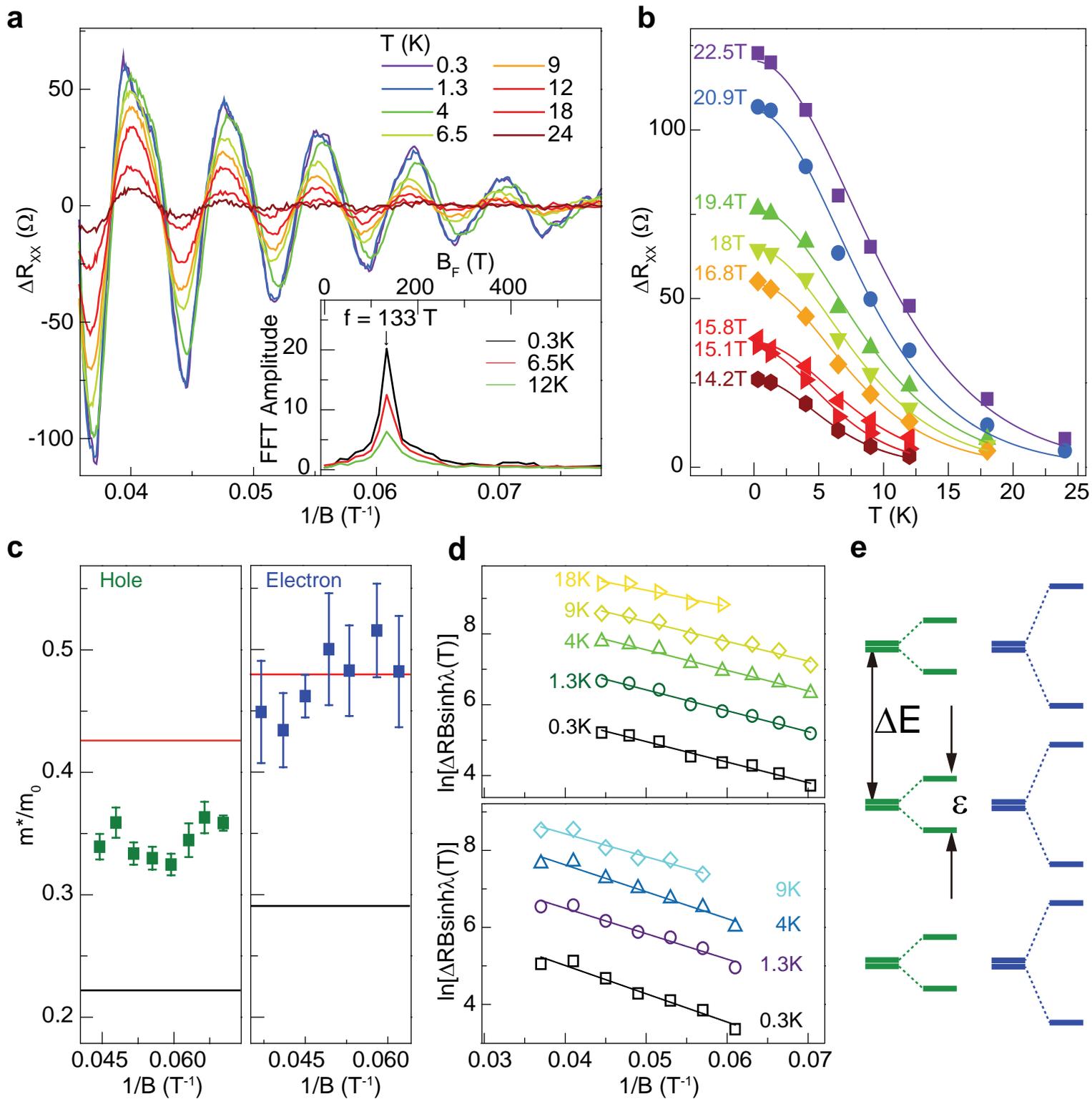

Likai Li *et al.* Figure 4

# Quantum Oscillations in Black Phosphorus Two-dimensional Electron Gas

Likai Li, Guo Jun Ye, Vy Tran, Ruixiang Fei, Guorui Chen, Huichao Wang, Jian Wang, Li Yang, Kenji Watanabe, Takashi Taniguchi, Xian Hui Chen[*] and Yuanbo Zhang[*]

*Email: zhyb@fudan.edu.cn, chenxh@ustc.edu.cn

# Supplementary Information

## 1. Modeling the surface charge accumulation in black phosphorus FET

We performed *ab initio* calculations, combined with model extrapolation, to determine the distribution of gate-induced charge carriers, and to extract the thickness of the 2DEG in our black phosphorus FET devices. The *ab initio* calculations were done with the SIESTA package, and we modeled our device with a 10-layer (~ 5.8 nm) black phosphorus slab that is doped to certain carrier density and subjected to gate electric field. We then obtained the free carrier distribution by integrating the local density of states between the Fermi level and the valence band maximum (conduction band minimum) for holes (electrons).

The *ab initio* calculations worked well up to a gate electric field of 0.4 V/nm (Fig. S1a, black curves), but higher fields close the black phosphorus band gap and render the calculations unmanageable. In order to get the charge distribution at fields up to 1.1 V/nm that are realized in our experiment, we analytically modeled the low-field distribution and extrapolated it to high fields. Specifically, we modeled the envelope of free carrier density with Airy function (Fig. S1a, red broken lines):

$$\varphi(z) = A \times Ai(\frac{z}{L} + a_1) \tag{S1}$$

where $a_1 \approx 2.338$ is the first zero of $Ai(z)$, and $A$ and $L$ are adjustable parameters. $\varphi(z)$ becomes exact if the confining potential is a triangular well. Here $A$ is determined by the total free charge and $L$ controls the width of the function. We extracted $A$ and $L$ by fitting $\varphi(z)$ with our *ab initio* calculations at low gate electric fields (Fig. S1a), and then extrapolated $A$ and $L$ to high fields (Fig. S1c-d). Our results show that most of the free carriers are confined within ~ 2 atomic layers (~ 1.1 nm) when the gate electric field is above 0.4 V/nm, confirming the 2D nature of our black phosphorus electron gas.

Finally, we calculated the subband energy levels in black phosphorus 2DEG induced by a gate electric field of 1.1 V/nm, the same field used in our experiment. We first obtained the confinement potential by solving Poisson's equation with the extrapolated charge distribution. The confinement potential, depicted in Fig. S1b (black lines), yields a subband splitting of ~ 300 meV for electrons (and ~ 250 meV for holes, not shown) as the solution to a one-dimensional Schrödinger equation. The lack of nodes in the charge density profile (Fig. S1c and S1d for electrons and holes, respectively), coupled to the fact that calculated density of surface states in the first subband is 2 orders of magnitude larger than the doping level experimentally observed in the sample, indicates that only the first subband is occupied in our experiment.

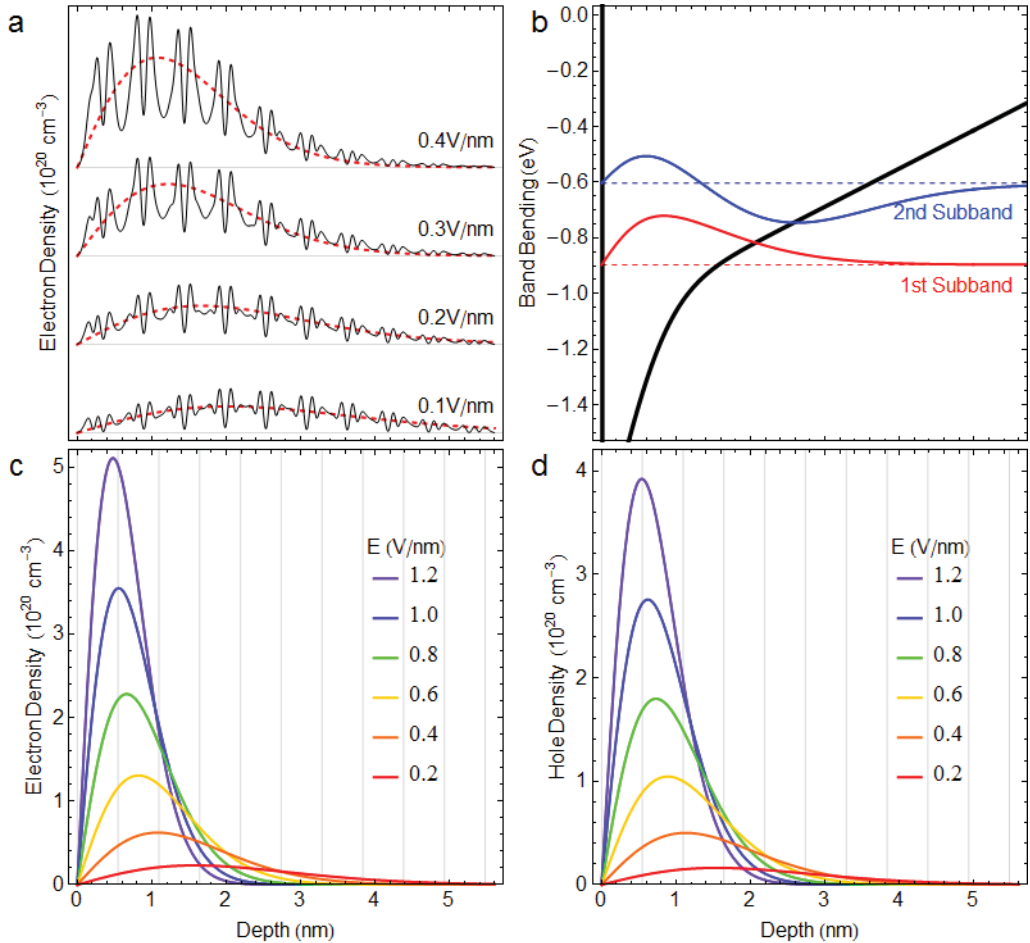

**Figure S1 | Distribution of gate-induced free carriers in black phosphorus. a,** *Ab initio* calculation of free carrier distribution (black) fitted with Airy envelope function defined in Eq. S1 (red dashed) for electrons. The free carriers are induced by varying gate-electric fields whose values are marked for each curve. **b,** Confinement potential (black) induced by a gate electric field of 1.1 V/nm. The wavefunction of lowest two subbands (red and blue) of electrons are also shown (arbitrary unit). **c and d,** Free carrier density as a function of depth measured from the surface calculated for electrons and holes, respectively. Carrier densities at high gate electric fields (> 0.4 V/nm) are obtained by extrapolation from the *ab initio* carrier density at low fields. Vertical lines mark the midpoints between phosphorene layers.

## 2. Crystal orientation determined by Raman spectroscopy

We performed Raman spectroscopy with linearly polarized light to determine the crystal orientation of our black phosphorus samples. The excitation laser ($\lambda = 532$ nm) was polarized in the *x-y* plane and incident along the *z* direction. As shown in Fig. S2c, three Raman peaks were observed, which corresponds to the $A_g^1$, $B_{2g}$, and $A_g^2$ vibration modes[S4,S5]. The $A_g^2$ peak intensity strongly depends on the polarization angle, while the other two modes do not show significant angular dependence (Fig. S2c). Since $A_g^2$ mode corresponds to lattice vibrations along *x* axis (perpendicular to zigzag direction, Fig. S2b), the $A_g^2$ peak intensity should be strongest when excitation laser is polarized along the *x* direction[S4,S6,S7]. Indeed, the angle-dependent $A_g^2$ peak intensity is well fitted with a cosine dependence (Fig. S2d), and we identified the *x* direction of the crystal as the direction on which the cosine function is maximum. The *x* direction of the crystal shown in Fig. 2a determined this way was found to be 21° away from the title axis in the angle-dependent magneto-resistance measurements (Fig. S3b. See also Fig. 3a)

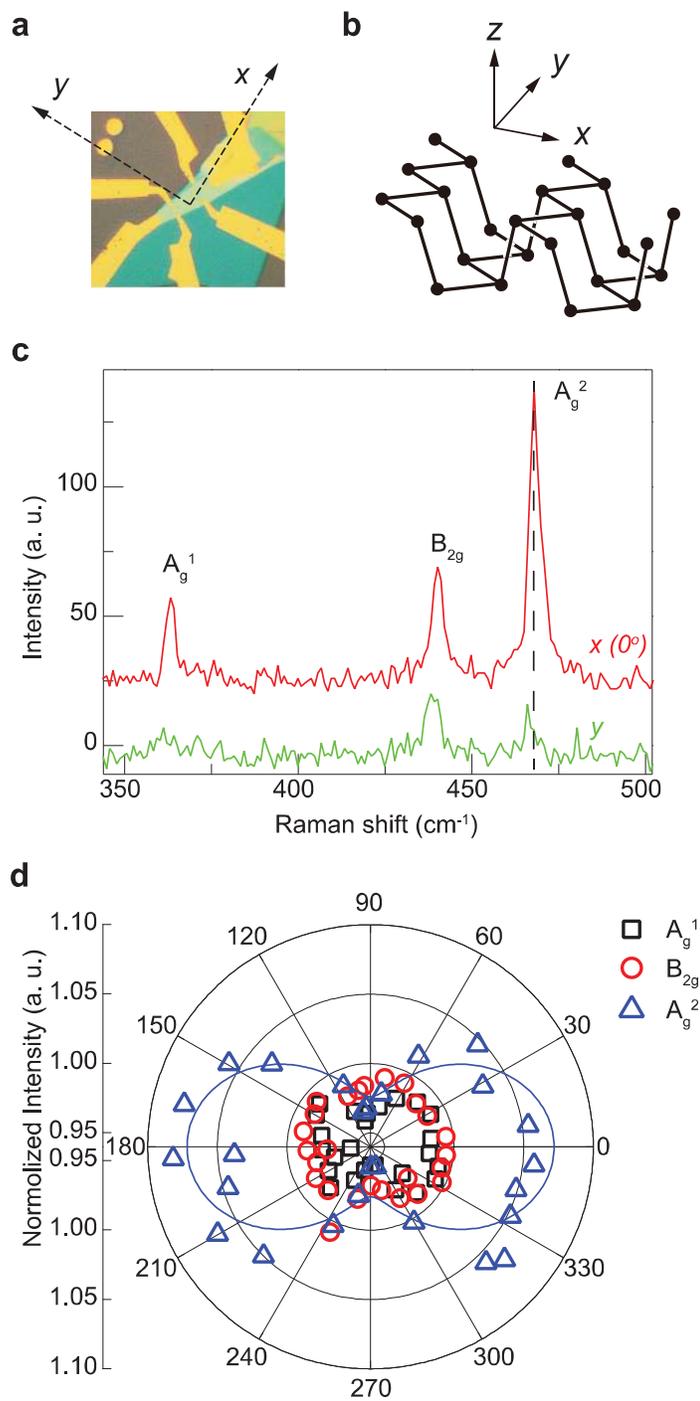

**Figure S2 | Determination of crystal orientation from angle-dependent Raman spectroscopy. a,** The same black phosphorus sample shown in Fig. 3a. The *x* and *y* axis denote the crystal orientation determined from angle-dependent Raman spectroscopy. **b,** Atomic structure of black phosphorus showing the *x*, *y* and *z* axis of the crystal. **c.** Raman spectra of the sample shown in **a** obtained with excitation laser polarized along *x* (red) and *y* (green) directions. The $A_g^2$ peak intensity shows pronounced variation at

these two polarization angles. **d.** Intensity of $A_g^1$, $B_{2g}$ and $A_g^2$ peaks plotted as a function of the polarization angle of the excitation laser. The intensity values are normalized to the Si peak from the substrate. The *x* axis of the crystal is determined by fitting the $A_g^2$ peak intensity by a cosine function (blue).

3. **Black phosphorus electron gas – 3D or 2D?**

In this section we will first calculate the angular dependence of the SdH oscillation frequency $B_F$ we would expect if our black phosphorus electron gas is 3D instead of 2D. We will then show that the 3D model of the electron gas does not fit our data, and our results can only be explained by a 2D black phosphorus electron gas.

Let us first assume the gate electric field induced electron gas in black phosphorus is 3D. In this case the Fermi surface can be approximated by a tri-axial ellipsoid:

$$\frac{2E_f}{\hbar^2} = \frac{k_x^2}{m_x^*} + \frac{k_y^2}{m_y^*} + \frac{k_z^2}{m_z^*} \tag{S2}$$

Here $m_x^* = 0.076 m_0$, $m_y^* = 0.648 m_0$ and $m_z^* = 0.28 m_0$ are taken from previous measurements of hole effective mass[S3] on bulk crystal; $E_f$ is the Fermi energy. The external cross-section of the Fermi surface $S_F$ seen by the external magnetic field (applied vertically in our case) determines the frequency of the SdH oscillations: $B_F = (2\pi)^2 e / h S_F$ (Ref. S8). As the sample is tilted about an in-plane axis (referred to as *i* axis shown in Fig. S3a) that is at an angle $\theta$ from the *x* axis, the external cross-section seen by the magnetic field becomes [S9]:

$$S_F = \left[ (\frac{\cos\alpha}{S_{F-xy}})^2 + (\frac{\sin\alpha}{S_{F-iz}})^2 \right]^{-\frac{1}{2}} \tag{S3}$$

Here $\alpha$ is tilt angle; $S_{F-xy} = \frac{2\pi E_f}{\hbar^2}\sqrt{m_x^* m_y^*}$ and $S_{F-iz} = \frac{2\pi E_f}{\hbar^2}\sqrt{m_z^* / [(\cos\theta)^2 / m_x^* + (\sin\theta)^2 / m_y^*]}$ are the external cross-sections in the *x-y* and *i-z* plane, respectively. $B_F$ as a function of $\alpha$ is therefore:

$$B_F = B_{F-xy} \left[ (\cos \alpha)^2 + (\sin \alpha \frac{S_{F-xy}}{S_{F-iz}})^2 \right]^{-\frac{1}{2}} \tag{S4}$$

where $B_{F-xy}$ is the oscillation frequency when the magnetic field is normal to the *x-y* plane. The normalized SdH oscillation frequency, $B_F / B_{F-xy}$, as a function of $1/\cos(\alpha)$ for different $\theta$ is plotted in Fig. S3c (solid lines). For a 2D electron gas, $B_F / B_{F-xy}$ is simply $1/\cos(\alpha)$, also plotted in Fig. S3c for comparison (black line). The vast difference between the expected angular dependence of $B_F / B_{F-xy}$ in the 3D and 2D case provides us an easy way to distinguish whether our black phosphorus electron gas is 3D or 2D.

We measured the $B_F / B_{F-xy}$ as a function of $1/\cos(\alpha)$, and the data are shown in Fig. S3c. We tilted our sample about two orthogonal axes ($\theta = -21°$ and $\theta = 69°$), and red circles and blue squares are observed angular dependence of $B_F / B_{F-xy}$ about each axis. Both data sets depart significantly from the dependence expected in the 3D case, and fall on the straight line corresponding to a 2D electron gas. Our results provide unambiguous evidence that our black phosphorus electron gas is 2D instead of 3D in nature.

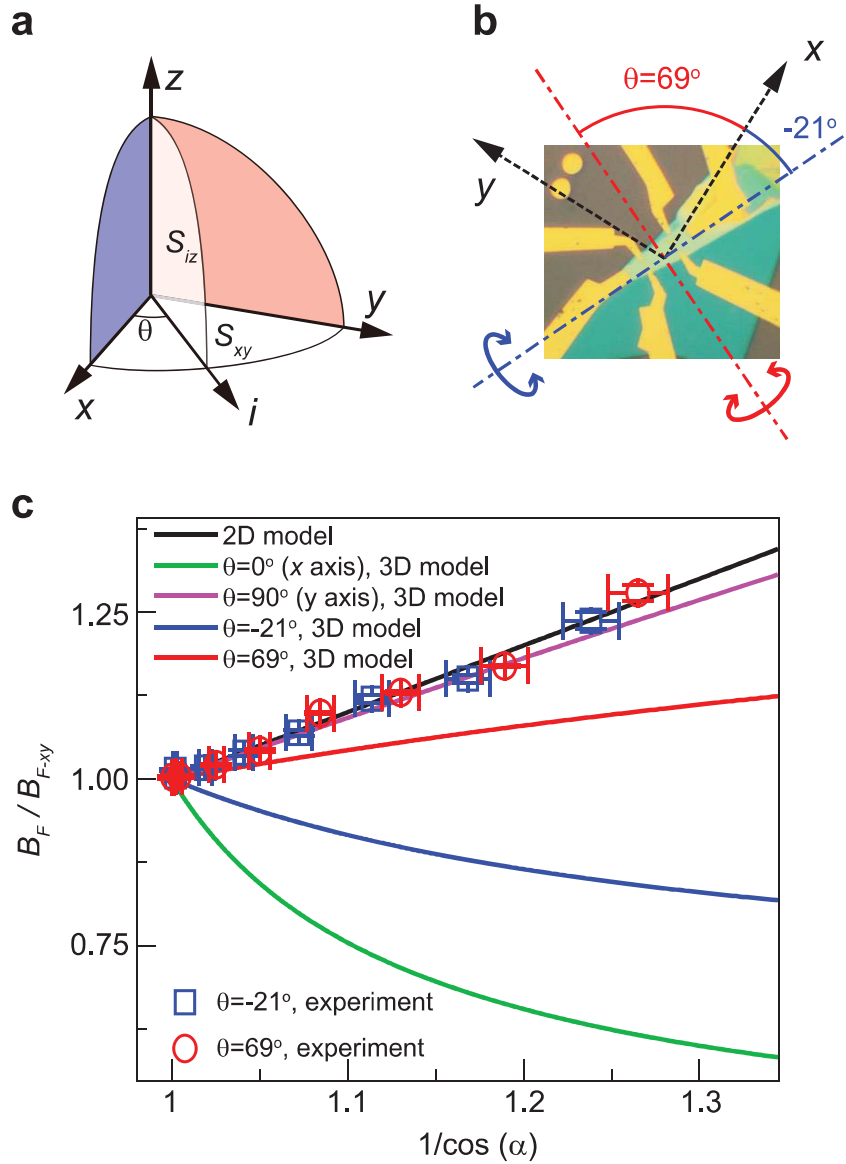

**Figure S3 | 2D nature of black phosphorus electron gas probed by the angular dependence of SdH oscillation frequency. a,** Schematic drawing of the Fermi surface of bulk black phosphorus. The *i* axis at an angle $\theta$ from the *x* axis represents the tilt axis in the *x-y* plane. **b,** Configuration of the angular dependence measurement. Two orthogonal tilt axes, $\theta = -21°$ and $\theta = 69°$, are shown in red and blue, respectively. *x* and *y* axes denote the crystal orientation of the sample. **c,** $B_F / B_{F-xy}$ as a function of $1/\cos(\alpha)$ as expected from model calculations (solid lines) and from our measurements (squares and circles). Red circles and blue squares are observed angular dependence as the sample is tilted about two orthogonal axes ($\theta = -21°$ and $\theta = 69°$). Both data sets depart significantly from the dependence expected in the 3D case (blue and red), and fall on the straight line (black) corresponding to a 2D electron gas.

## 4. Measurement of cyclotron mass of electrons

In the main text, we have obtained the hole cycltron mass by fitting the temperature-dependent SdH oscillation amplitude with the reduction factor $R_T$. This procedure works well for holes ($V_g < 0$), because the SdH oscillations contain only a single frequency (Fig. 4a), and the data are well described by Eq. (1). Problem arises if we try to use the same procedure to extract the cyclotron mass on the electron side of the doping ($V_g > 0$). The Zeeman splitting of the LLs introduces the second harmonic into the SdH oscillations (Fig. S4a), and the thermal damping of the oscillation amplitude is no longer described by $R_T$.

However, an accurate determination of the electron cyclotron mass is still possible if the higher harmonics are taken into account. Since the oscillatory part of $R_{xx}$ is dominated by the first and second harmonics on the electron side (Fig. S4a, inset), the oscillations are described by[S8]

$$\Delta R_{xx} = \Delta R_0 \sum_{p=1,2} c_p \frac{p\lambda(T)}{\sinh[p\lambda(T)]} e^{-pD} \cos(p\phi) \cos[2\pi p(B_F/B + 1/2)] \quad (S5)$$

Here $c_{1(2)}$ is the coefficient of the first (second) harmonic, and $\lambda(T)$, $D$ and $\phi$ are defined in the main text. Indeed, by simply adding the second harmonic component to a 'simulated' oscillation signal, we can reproduce the main features in our data (Fig. S5b). As the main peaks are now split into two and minor valleys develop between the split peaks, the definition of the 'amplitude' of the SdH oscillations needs to be clarified. We define the amplitude as the difference in $R_{xx}$ between a major valley and its adjacent minor valley. As is clear in Fig. S4b, the amplitude by this definition comprises contributions from both harmonics: the first harmonic contributes a $\delta R_{xx}$ between a peak and valley; and the second harmonic contributes a $\delta R_{xx}$ between a valley and its neighbouring valley (or a peak and its neighbouring peak). Both terms depend on temperature as prescribed by Eq. (S5), and the temperature reduction factor becomes:

$$R_T = \Delta R_0 \frac{2\pi^2 k_B T m^*/\hbar eB}{\sinh(2\pi^2 k_B T m^*/\hbar eB)} + \Delta R_0' \frac{4\pi^2 k_B T m^*/\hbar eB}{\sinh(4\pi^2 k_B T m^*/\hbar eB)} \quad (S6)$$

where $\Delta R_0$ and $\Delta R_0'$ are temperature-independent free parameters. By fitting our data with Eq. (S6), we extracted the cycltron mass of electrons (Fig. S4c, dashed lines), and the results are shown in Fig. S4d (red filled circles). Here fitting with both harmonics is only possible for $B > 20\,\text{T}$, where the Zeeman splitting of the LLs are clearly visible. We also carried out the fitting with only the first harmonic at $B > 20\,\text{T}$ (Fig. S4c, solid lines), and the resulting cyclotron masses do not show appreciable deviation. This result indicates that $R_T$ is dominated by the first harmonic term in Eq. (S6). At low fields ($B < 20\,\text{T}$), Zeeman splitting becomes negligible, and fitting with only the first harmonic is sufficient (Fig. S4d, black squares). In the main text, we displayed the electron cyclotron mass at both low fields ($B < 20\,\text{T}$) and high fields ($B > 20\,\text{T}$), obtained from fitting with only first harmonic and with first and second harmonics, respectively.

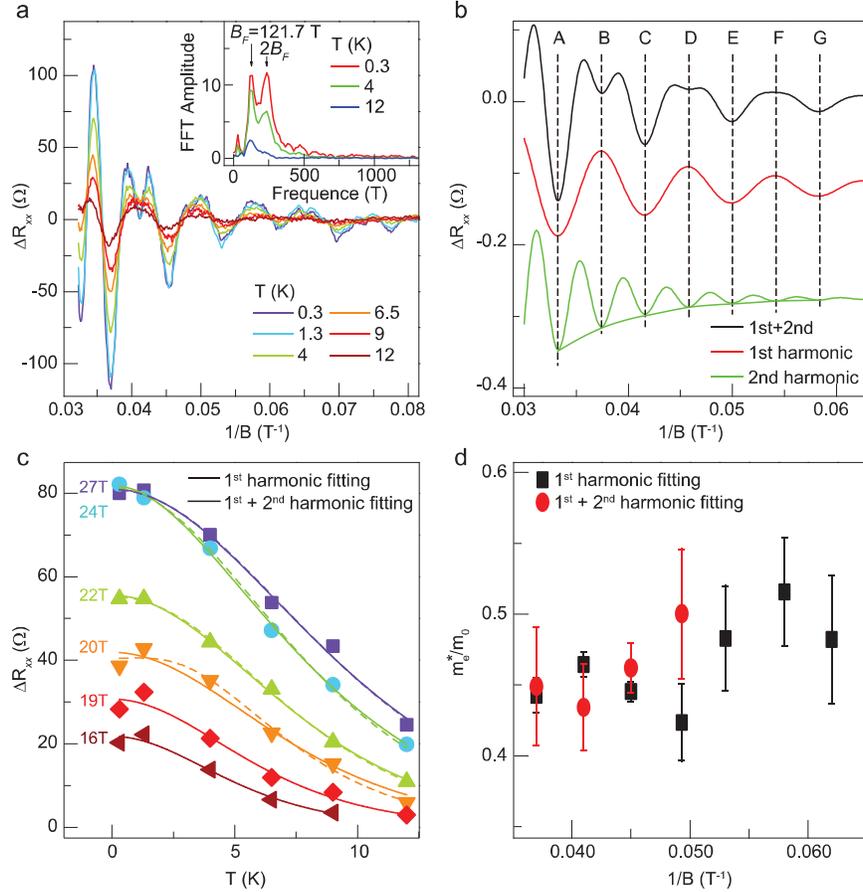

**Figure S4 | Measurement of cyclotron mass of electrons. a,** $\Delta R_{xx}$ as a function of $1/B$ at different temperatures at $V_g = +100$ V. Pronounced Zeeman splitting is observed at low temperatures. Inset: Fast Fourier transformation (FFT) of the oscillations shown in the main panel. The oscillations are dominated by the first and second harmonic components. **b.** Simulated SdH oscillations with only first and second harmonics, which reproduce the main features observed in **a**. **c.** SdH oscillation amplitude extracted from **a** as a function of temperature at different magnetic fields. Broken lines are fit with Eq. (S6), and solid lines are fit with only first harmonic taken into account. **d.** Electron cyclotron mass obtained from the fit in **c**.

## Calculating the effective mass of electrons and holes

We used first-principle density functional theory (DFT) calculation with the PBE-van-der Waals functional to obtain the fully relaxed atomic structure of black phosphorus. The energy band structure was then calculated using the HSE06 hybrid functional. A plane-wave basis was used in all our calculations with a 45 Ry energy cutoff and a norm-conserving pseudopotential. The sampling grid in $k$ space was $14 \times 10 \times 1$ for few-layer structures and $14 \times 10 \times 4$ for the bulk. The effective mass ($m_x$ along the armchair and $m_y$ along the zigzag direction) of free carriers were obtained by fitting the band-edge dispersion near the Fermi level with a parabola (the Fermi level is determined from the gate-induced doping level used in the experiment). So the hole masses were calculated at energies slightly below the VBM, and the electron masses were at energies slightly above the CBM. $m_x$, $m_y$, and cyclotron mass $m^* = \sqrt{m_x m_y}$ are shown in Table S1 and S2 in unit of bare electron mass.

Table S1: Effective mass of electrons

| Number of layers | $m_x$ | $m_y$ | $m^*$ |
| --- | --- | --- | --- |
| 1 | 0.146 | 1.246 | 0.427 |
| 2 | 0.136 | 1.334 | 0.426 |
| Bulk | 0.140 | 0.710 | 0.307 |

Table S2: Effective mass of holes

| Number of layers | $m_x$ | $m_y$ | $m^*$ |
| --- | --- | --- | --- |
| 1 | 0.146 | 6.560 | 0.979 |
| 2 | 0.143 | 1.610 | 0.480 |
| Bulk | 0.122 | 0.720 | 0.296 |